\documentclass[10pt,letterpaper,twocolumn]{article} 

\usepackage{ol2}
\usepackage{amsmath}
\usepackage{hyperref,cite}
\usepackage{color}

\newcommand{\Fig}[1]{Figure~\ref{#1}}

\newcommand{\fig}[1]{Fig.~\ref{#1}}

\newcommand{\mic}{~{\mu \rm m}}

\begin{document}

\twocolumn[ 
\title{Efficient supercontinuum generation in quadratic nonlinear waveguides without quasi-phase matching}
\vspace{-3 mm}
\author{Hairun Guo$^1$, Binbin Zhou$^1$, Michael Steinert$^2$, Frank Setzpfandt$^2$, Thomas Pertsch$^2$, Hung-ping Chung$^3$, Yen-Hung Chen$^3$,
and Morten Bache$^{1,*}$}
\address{
$^1$DTU Fotonik, Department of Photonics Engineering, Technical University of Denmark, DK-2800 Kgs. Lyngby, Denmark \\
$^2$Institute of Applied Physics, Abbe Center of Photonics, Friedrich-Schiller-Universit{\"a}t Jena, 07743 Jena, Germany\\
$^3$Department of Optics and Photonics,
National Central University, Jhongli 320, Taiwan\\
$*$\href{mailto:moba@fotonik.dtu.dk}{moba@fotonik.dtu.dk} \\
}
\vspace{-3 mm}

\begin{abstract}
  Efficient supercontinuum generation (SCG) requires excitation of solitons at the pump laser wavelength. Quadratic nonlinear waveguides may support an effective self-defocusing nonlinearity so solitons can directly be generated at common ultrafast laser wavelengths without any waveguide dispersion engineering. We here experimentally demonstrate efficient SCG in a standard lithium niobate (LN) waveguide without using quasi-phase matching (QPM). By using femtosecond pumps with wavelengths in the $1.25-1.5\mic$ range, where LN has normal dispersion and thus supports self-defocusing solitons, octave-spanning SCG is observed. An optimized mid-IR waveguide design is expected to support even broader spectra. The QPM-free design reduces production complexity, allows longer waveguides, limits undesired spectral resonances and effectively allows using nonlinear crystals where QPM is inefficient or impossible. This result is important for mid-IR SCG, where QPM-free self-defocusing waveguides in common mid-IR nonlinear crystals can support solitons directly at mid-IR ultrafast laser wavelengths, where these waveguides have normal dispersion.
\end{abstract}
] 
\vspace{-6 mm}

\noindent 
Supercontinuum generation (SCG) using stable ultrafast pulsed lasers in fibers and waveguides has found a wide range of applications in fields such as frequency metrology, optical coherence tomography and spectroscopy \cite{Dudley:2010}.
Especially soliton-induced SCG is efficient, which 
relies on balancing the material self-focusing Kerr nonlinearity with anomalous dispersion, i.e. pumping above the fiber zero-dispersion wavelength (ZDW). 
This obstacle was overcome in the near-IR through advanced silica fiber production technology, which allowed creating sophisticated photonic crystal fibers (PCFs) with strong waveguide dispersion so the waveguide ZDW could be shifted down to operating wavelengths of common ultrafast pulsed lasers.
Current efforts target efficient and ultrafast mid-IR SCG sources \cite{Yu2013}, mainly motivated by the so-called molecular vibrational "fingerprint region" in the mid-IR. As mid-IR transparent glasses have ZDW well into the mid-IR, far from pump wavelengths of emerging mid-IR ultrafast lasers, a strong waveguide dispersion is needed here as well. However, to date only simple PCF designs have been made \cite{Domachuk2008}, and the most efficient fiber-based mid-IR SCG to date used a pump wavelength chosen to match the fiber dispersion \cite{Petersen:2014}.

In contrast, for a self-defocusing nonlinearity soliton formation requires normal dispersion, i.e. pumping below the ZDW, so a self-defocusing waveguide is free from waveguide dispersion constraints. Uniquely, this is possible using cascaded nonlinearities in quadratic nonlinear crystals, where the crystal's intrinsic self-focusing nonlinearity ($n_{\rm 2,Kerr}>0$) is counterbalanced by a self-defocusing cascading nonlinearity $n_{\rm 2,casc}<0$, leading to an effective self-defocusing effect $n_{2,\rm eff}\equiv n_{\rm 2,casc}+n_{\rm 2,Kerr}<0$. In bulk this has found numerous applications, e.g. for pulse compression and temporal soliton formation \cite{liu:1999,ashihara2002soliton,moses2006soliton,zhou2012ultrafast} and SCG \cite{fuji2005monolithic,zhou2012ultrafast,Zhou:2014a,Zhou2014}. Motivated by the promise of efficient soliton-induced SCG, self-defocusing waveguides were pumped with ultrafast near-IR fiber lasers in the normal dispersion regime \cite{langrock2007generation,Phillips:2011-ol}: the material ZDW of the waveguide material (lithium niobate, LN) is as high as $1.9\mic$. As the near-IR pumps could directly excite solitons, efficient broadband SCG was observed.

The most efficient cascaded nonlinearity is phase-mismatched second-harmonic generation (SHG) where the induced self-defocusing nonlinearity is $n_{2,\rm casc}\propto -d_{\rm eff}^2/\Delta k$ \cite{desalvo1992self}. Here $d_{\rm eff}$ is the effective quadratic nonlinearity and $\Delta k=k_{2\omega_1}-2k_{\omega_1}$ is the SHG phase-mismatch. 
Quadratic nonlinear waveguides usually have a large $\Delta k$. Consequently quasi-phase-matching (QPM) is used to reduce $\Delta k$ and thereby enhance cascading \cite{sundheimer1993large}; also in \cite{langrock2007generation,Phillips:2011-ol} QPM was used. However, the penalty of QPM is that $d_{\rm eff}$ is reduced, so to increase cascading requires a significant reduction in $\Delta k$. This in turn increases the cascading-induced pulse self-steepening ($\propto 1/\Delta k$) \cite{moses2006controllable} and may lead to resonances in the cascaded nonlinear "response" \cite{bache2007nonlocal,bache2008limits,zhou2012ultrafast}. This is detrimental to ultrashort pulse interaction, which requires an ultra-broadband nonlinearity. Additionally, using QPM for cascading implies that no frequency conversion process are phase matched, and this in turn results in undesired resonant phase-matching peaks in the spectrum, especially from higher-order QPM interactions. A QPM-free waveguide is also much less complex to make and can be made very long. Finally, whether QPM can be used for the many mid-IR nonlinear crystals that promise an effective self-defocusing nonlinearity \cite{Bache:2013-midIR} remains to be proven.

We aim here to show that a QPM-free quadratic nonlinear waveguide can be self-defocusing and that efficient soliton-induced SCG is possible when pumping in the normal dispersion regime. As a first demonstration of this principle we choose a buried-core LN waveguide pumped with a femtosecond near-IR laser. While the generated bandwidths are similar to previous results \cite{langrock2007generation,Phillips:2011-ol}, we expect that a much broader supercontinuum can be generated by optimizing the waveguide. 
However, the main attraction is that this promises well for realizing QPM-free self-defocusing waveguides in mid-IR crystals, which would be excellent candidates for a soliton-based mid-IR SCG source since their self-defocusing nonlinearity will naturally support solitons at the wavelengths of emerging ultrafast mid-IR lasers.

The idea behind a QPM-free self-defocusing cascaded interaction in LN cut for type-0 noncritical ($ee\rightarrow e$) interaction was first suggested and observed in a bulk crystal \cite{zhou2012ultrafast}, where both soliton self-compression and octave-spanning SCG were observed when pumping around $1.3\mic$, and was later extended to the range from $1.2-1.45\mic$ \cite{Zhou:2014a}. Theoretically, solitons can be excited with pump wavelengths as high as the ZDW of LN ($1.9\mic$), where effective self-defocusing nonlinearity should be even higher due to a reduced $\Delta k$ \cite{zhou2012ultrafast}. We can therefore expect that a near-IR pumped QPM-free LN waveguide will support self-defocusing solitons provided that the waveguide dispersion is small. Recently a QPM-free LN ridge waveguide design was proposed \cite{guo2014few}, where the LN waveguide had a low refractive-index (RI) contrast so the dispersion profile remained similar to that of bulk LN. In turn, the waveguide did not alter much the nonlinear balance, effectively giving a similar cascading as the bulk case, namely self-defocusing over a broad wavelength range where the dispersion was normal. 

We here attempt to implement these ideas in low-RI contrast LN buried-core waveguides fabricated by annealed proton exchange (APE) on a z-cut congruent LN wafer, cut for type-0 noncritical $ee\rightarrow e$ SHG ($\theta=\pi/2$, $|\phi|=\pi/2$). The waveguides were 5.0 mm long, had a depth of $4.1 ~{\rm \mu m}$ and four different widths were fabricated (between $6.0-9.0\mic$ with $1.0\mic$ intervals). The RI contrast was $\Delta n = 0.029$, measured at 1550 nm, which is comparable to the waveguide in \cite{guo2014few}. The low RI contrast will ensure single-mode guidance in near-IR.

In order to assess whether the waveguides are self-defocusing, consider the effective waveguide nonlinearity responsible for self-phase modulation (SPM) is $\gamma_{\rm eff} = \gamma_{\rm casc}+\gamma_{\rm Kerr}$; the cascaded nonlinearity is $\gamma_{\rm casc}=\tfrac{\omega_1}{c}n_{2,\rm casc}/A_{\rm eff,SHG} \propto -d_{\rm 33}^2/[\Delta k A_{\rm eff,SHG}]$ and $\gamma_{\rm Kerr}= \tfrac{\omega_1}{c}(1-f_R)n_{2,\rm Kerr}/A_{\rm eff,SPM}\propto (1-f_R)c_{\rm 33}/A_{\rm eff,SPM}$  is the material self-focusing electronic Kerr nonlinearity and $f_R$ is the Raman fraction \cite{guo2014few}. We have used that in the noncritical type-0 interaction the effective quadratic and cubic nonlinearities are $d_{33}$ and $c_{33}$, respectively \cite{guo2013nonlinear}. $A_{\rm eff,SHG}$ and ${A_{\rm eff,SPM}}$ are effective mode areas corresponding to quadratic SHG and cubic SPM processes, and usually $A_{\rm eff,SHG} \approx A_{\rm eff,SPM}$. For this reason, an estimate of whether the QPM-free waveguide is self-defocusing can be made from the bulk nonlinear coefficients $n_{2,\rm eff}\equiv n_{\rm 2,casc}+n_{\rm 2,Kerr}$, provided that the waveguide does not change the bulk $\Delta k$-value significantly.  
Due to the low RI contrast, $\Delta k$ is not altered much and we can therefore be quite certain that the waveguides are effectively self-defocusing. Extending this to the mid-IR implies that the cascading figure-of-merit of bulk crystals \cite[Fig. 1]{Bache:2013-midIR} can be used as a good indication of potential crystals for mid-IR self-defocusing waveguides.

The experiment used a commercial regenerative amplifier (1 kHz repetition rate) followed by an optical parametric amplifier, giving pulses with $\tau _{\rm FWHM} = 50 ~{\rm fs}$ (full-width half-maximum) duration that were tuned in the range $1250-1500 ~{\rm nm}$.
The beam was spatially coupled into the waveguide by an objective lens (Nikon $\times 20$/0.5) and the output was collected by an aspheric lens (Thorlabs, $f = 8.0 ~{\rm mm}$, $NA = 0.5$, near-IR AR coated). We estimate a 10 dB coupling loss, mainly due to interface reflection on lenses as well as on waveguide ends, and from  mode-mismatch between the input beam and the waveguide fundamental $\rm TM_{00}$ eigenmode. By measuring the average power after the waveguide, the maximum pulse energy inside the waveguide was estimated to 10 nJ, i.e. a peak power of $200 ~{\rm kW}$. 
The output pulse spectrum was measured by a spectrometer (OceanOptics NIRQuest256-2.5) via a collecting fiber, which means that the center position of the output was recorded.

\begin{figure}[t]
  \centering{
  \includegraphics[width = 1 \linewidth]{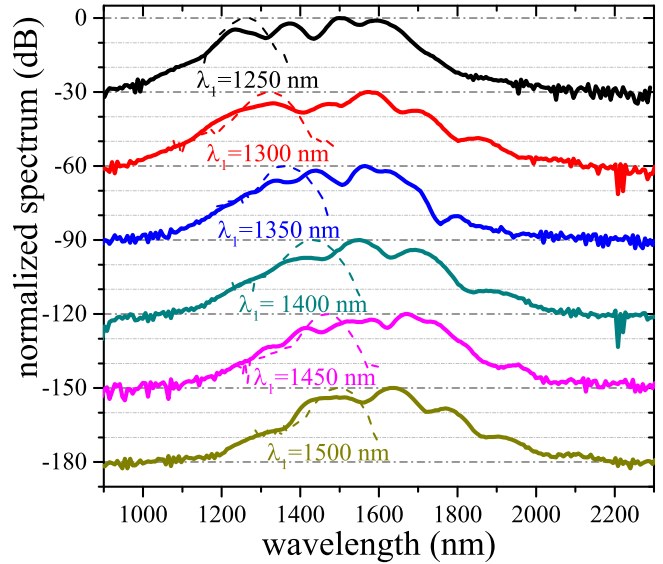}
  }
  \vspace{-3 mm}
  \caption{
  SCG in a 5 mm LN APE QPM-free waveguide having $9.0\mic$ width using different pump wavelengths. The output spectra (solid lines) are normalized and recorded for $P_{\rm peak} \approx 200 ~{\rm k W}$, while the input spectra (dashed lines) were recorded at low power. For clarity, a 30-dB offset per spectrum is used.
  }
  \vspace{-3 mm}
  \label{fig-2}
\end{figure}

\Fig{fig-2} shows SCG in the $9.0\mic$ wide waveguide by using different pump wavelengths from 1250 nm to 1500 nm. These spectra were recorded with a pulse peak power of 200 kW and energy of 10 nJ.
The spectral broadening could easily exceed 600 nm (at $-20$ dB) and it becomes octave-spanning (from $1 \sim 2 ~{\rm \mu m}$) at $-30$ db. This bandwidth is similar to \cite{langrock2007generation} even if our waveguide was over 6 times shorter. On the other hand, quite high peak powers (around 80 kW and above, see more later) are required to achieve such a broad bandwidth due to the limited confinement of the large core. The spectra show a significant asymmetry to longer wavelengths, which is mainly caused by red-shifting of the soliton by the Raman effect of LN (see also, e.g., \cite{phillips2011supercontinuum-oe,zhou2012ultrafast,bache2011optical,guo2013nonlinear}). 

\begin{figure}[t]
  \centering{
  \includegraphics[width = 1 \linewidth]{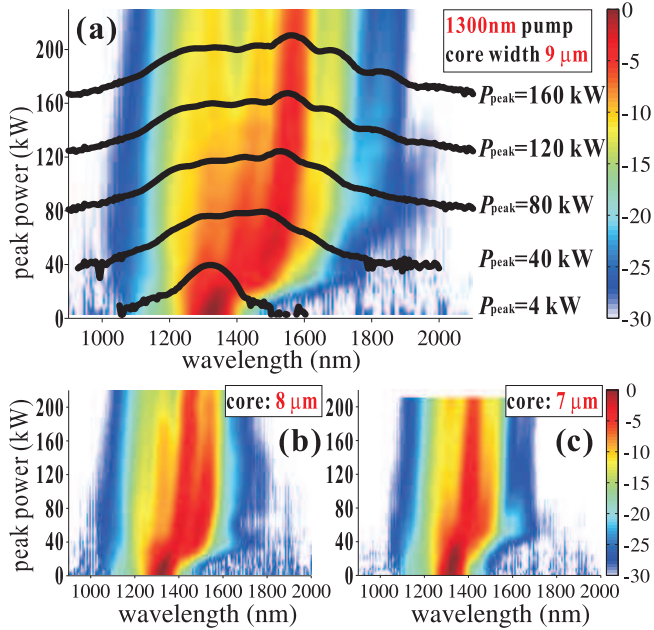}
  }
  \vspace{-3 mm}
  \caption{
Pulse spectral evolution with respect to the average pulse power. The waveguide core widths are (a) $9 ~{\rm \mu m}$, (b) $8 ~{\rm \mu m}$ and (c) $7 ~{\rm \mu m}$. Note that each spectrum is normalized to the corresponding pulse energy.  }
  \vspace{-3 mm}
  \label{fig-3}
\end{figure}

\begin{figure}[t]
  \centering{
  \includegraphics[width = 1 \linewidth]{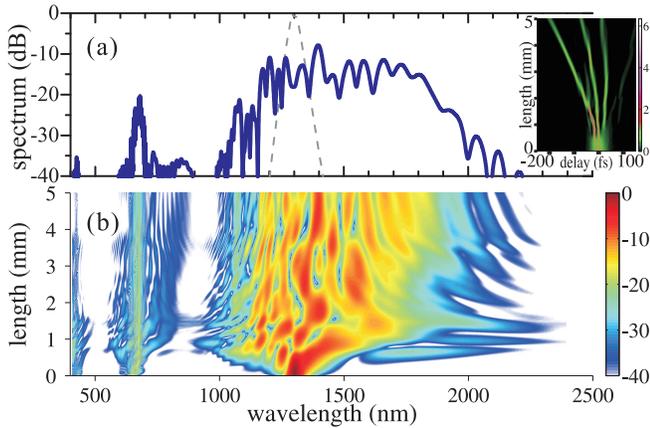}
  }
  \vspace{-3 mm}
  \caption{
  Numerical simulation of the $9.0\mic$ wide waveguide. (a) output spectrum; (b) pulse spectral evolution during the propagation; inset in (a) pulse temporal evolution during the propagation. The pump pulse has $\tau_{\rm FWHM}= 50 ~{\rm fs}$, $P_{\rm peak}= 120 ~{\rm kW}$, giving $\gamma_{\rm casc}= -47.4 ~{\rm (km \cdot W)^{-1}}$, $\gamma_{\rm Kerr}= 38.8 ~{\rm (km \cdot W)^{-1}}$, $L_{\rm D}(\lambda = 1.3\mic)= 4.63 ~{\rm mm}$, $L_{\rm N}= 0.97 ~{\rm mm}$. The simulations used $d_{33}=23.5$ pm/V and $(1-f_R)c_{33}=3.6\cdot 10^{-21}~\rm m^2/V^2$ (calculated by the 2-band model using $E_g=3.9$ eV), see \cite{Bache:2013-midIR,guo2013nonlinear} for details, and include Raman effects with $f_{\rm R} = 55 \%$ \cite{guo2013nonlinear}. 
  }
  \vspace{-3 mm}
  \label{fig-5}
\end{figure}

\Fig{fig-3} shows how the spectral broadening changes while sweeping the peak power. Generally, at low peak powers the bandwidth expands slowly. This is followed by a stage with rapid expansion of the bandwidth ($20<P_{\rm peak}<80~\rm kW$), after which a saturation occurs.
In order to understand this we remind that in a typical SCG process, the pulse spectral broadening is initiated by SPM effects: the bandwidth for a fixed waveguide length $L$ evolves as $\Delta \omega \propto \frac{\gamma_{\rm eff}P_{\rm peak}L}{\tau _{\rm FWHM}}$ \cite{agrawal2013nonlinear}. This expression implicitly assumes an unchanged temporal profile (no dispersion), and is therefore only accurate in the initial propagation stage at high peak powers. More quantitatively, SPM dynamics dominates when the pulse dispersion length $L_{\rm D}$ and the nonlinearity length $L_{\rm N} = (|\gamma_{\rm eff}|P_{\rm peak})^{-1}$ have the relationship: $L_{\rm D} \gg L_{\rm N}$. For the experiment in \Fig{fig-3}(a) we calculate $\gamma_{\rm eff}=-8.6~({\rm km\cdot W})^{-1}$, see also \Fig{fig-5}, and therefore estimate that $L_{\rm D} \simeq L_{\rm N}$ at around $P_{\rm peak}=20~\rm kW$. Thus, below this peak power $L_{\rm D} < L_{\rm N}$ and SPM spectral broadening is weak as dispersion dominates, while above $L_{\rm D} > L_{\rm N}$, which is an intermediate range where both dispersion and SPM affects the pulse. This initially leads to significant broadening and with an increased peak power there will be a gradual transformation into the soliton stage. The critical peak power where a soliton will self-compress inside the device length $L$ can be estimated from the scaling laws of soliton compression \cite{bache:2007} through the effective soliton order $N_{\rm eff}=\sqrt{L_{\rm D} / L_{\rm N}}$: around $P_{\rm peak}=80~\rm kW$ we find that the self-compression point $z_{\rm opt}$ is for the first time shorter than the 5 mm waveguide length. The self-compression of the soliton is naturally accompanied by a massive spectral broadening.

Following this stage, the increasing peak power means that $z_{\rm opt}<L$, i.e. that soliton self-compression occurs inside the waveguide. This means we should expect an accompanying mid-IR Cherenkov dispersive wave from a resonant phase-matching condition to the soliton due to higher-order dispersion \cite{bache:2010e,bache2011optical,Zhou:2014a}. However, we were unable to observe any significant mid-IR radiation, which we attribute to the poor mode confinement for $\lambda>2.0\mic$. At this stage soliton fission will also occur, mainly caused by Raman effects. Unfortunately, as these fissioned solitons will not induce mid-IR dispersive waves either, they will not contribute significantly to the spectral broadening. We finally mention that due to the low average power ($\mu$W level) and low repetition rate, we could not temporally characterize the output pulses. We hope soon to solve this issue by using a femtosecond ultrafast (MHz repetition rate) laser.

We can compare the spectral evolutions with respect to the peak power in differently sized waveguides in \fig{fig-3}(a-c). One evident trend is that by decreasing the waveguide width, the pulse spectral broadening is degraded, which implies a reduced $\gamma_{\rm eff}$. This seems counterintuitive as a smaller core should provide stronger confinement and thus higher nonlinearity (through smaller effective mode areas). However, ${\Delta k}$ increases as well due to the stronger confinement. Thus, the cascaded nonlinear strength $|{\gamma_{\rm casc}}|$ will not increase as fast as ${\gamma_{\rm Kerr}}$, and therefore the effective nonlinearity $\gamma_{\rm eff} = \gamma_{\rm casc}+\gamma_{\rm Kerr}$ drops. This fact also evidences the dilemma of using strong confinement \cite{Guo:2014}: eventually $\gamma_{\rm eff}$ becomes self-focusing, and one has to use QPM to regain self-defocusing. We finally mention that using the $6.0\mic$ wide waveguide minimal spectral broadening was observed and only for high peak powers and long pump wavelengths: clearly in this case $\gamma_{\rm eff}\simeq 0$.

We also performed numerical simulations of the APE waveguides: the waveguide core was assumed rectangular with a step-index change to the LN substrate; this gave qualitatively the same results as a realistic, but certainly more ambiguous, smooth APE RI profile. Then, we followed the steps in \cite{guo2014few} to: 1) calculate the waveguide eigenmodes, including the mode effective RI and the mode distribution; 2) estimate both cascaded and Kerr nonlinearities in the waveguide, including the calculation of effective mode areas; and 3) investigate the pulse propagation dynamics and ultra-broadband nonlinear interactions by solving a generalized nonlinear wave equation in frequency domain \cite{guo2013nonlinear}. The calculations showed that the TM$_{00}$ waveguide mode has normal dispersion below $2.0\mic$ and that the waveguide was effectively self-defocusing beyond $\lambda_1=1.05\mic$ \cite{Guo2014-CLEO}. \Fig{fig-5} shows a typical simulation with a 50 fs 1300-nm pump, $P_{\rm peak} = 120 ~{\rm kW}$. The output spectrum shows good qualitative agreement to the experimental data. It also evidences a weak SH component between $600-900$ nm; this was also observed experimentally, but was not included in Figs. \ref{fig-2}-\ref{fig-3} as it was recorded with a different spectrometer that was not calibrated to the near-IR spectrometer. In the simulation we can study the pulse temporal evolution in detail. It shows an SPM broadening stage leading to the soliton self-compression phase after 1 mm followed by soliton fission, induced by the Raman effect of LN and perturbations from higher-order dispersion. Notably, neither of the solitons emit mid-IR Cherenkov dispersive waves, and the spectrum does not extend beyond $2.2\mic$, quite similar to the experimental data.
We attribute this to the poor core confinement of the TM$_{00}$ mode beyond $2.0\mic$, which suppresses the dispersive wave emission.

Concluding we experimentally demonstrated SCG in QPM-free quadratic nonlinear LN waveguides designed for strongly phase-mismatched cascaded SHG. The broad spectra were observed for a range of pump wavelengths and peak powers. The results indicated that under weak waveguide confinement (large core size) the waveguide nonlinearity was similar to that of bulk LN, i.e. effectively self-defocusing even without QPM. In turn a high peak power ($\simeq 100$ kW) was needed to achieve maximal broadening.
The presented results are significant because self-defocusing nonlinear waveguides support solitons when pumping in the normal dispersion regime, and the QPM-free design allows using mid-IR crystals where normal dispersion is found at lasing wavelengths of  current or future ultrafast mid-IR lasers. This removes the strong dispersion engineering constraints that plague self-focusing fibers and waveguides, and which currently poses a significant challenge for efficient mid-IR SCG. Self-defocusing QPM-free waveguides in mid-IR crystals therefore present a radically new solution to achieve efficient and bright mid-IR supercontinua.

M.B. and B.Z. acknowledge the Danish Council for Independent Research (274-08-0479, 11-106702) for support. M.S., F.S., and T.P. acknowledge the German Research Foundation DFG (SPP1391 Ultrafast Nanooptics) for support.

\end{document}